\documentclass[preprint]{iacrtrans}
\usepackage[utf8]{inputenc}
\usepackage{amsmath}
\usepackage{caption}
\usepackage{refstyle}
\usepackage{cite}
\usepackage{adjustbox}
\usepackage{lipsum}% example text
\usepackage{graphics}
\usepackage{longtable}
\usepackage{makecell}
\usepackage{rotating} 
\usepackage{pdflscape} % for 'landscape' environment
\usepackage{graphicx}

\author{Ayoub Mars and  Wael Adi}
\institute{IDA, Institute of Computer and Network Engineering \\
Technical University of Braunschweig, Germany\\ 
\email{a.mars@tu-bs.de,  w.adi@tu-bs.de}}
\title{New Family of Stream Ciphers as Physically Clone-Resistant VLSI-Structures\footnote{This paper is under submission, and uploaded here only for comments and suggestions, and not for any commercial use.}}

\begin{document}

\maketitle

%%KEYWORDS
\keywords{Stream Cipher\and keystream generator\and NLFSR\and linear complexity\and Secret Unknown Cipher\and Physical Unclonable Functions\and Self-reconfigurating SoC FPGAs}

%%ABSTRACT
\begin{abstract}
A new large class of $2^{100}$ possible stream ciphers as key stream generators KSGs, is presented. The sample cipher-structure-concept is based on randomly selecting a set of 16 maximum-period Nonlinear Feedback Shift Registers (NLFSRs). A non-linear combining function is merging the 16 selected sequences. All resulting stream ciphers with a total state-size of 223 bits are designed to result with the same security level and have a linear complexity exceeding $2^{81}$ and a period exceeding $2^{161}$. A Secret Unknown Cipher (SUC) is created randomly by selecting one cipher from that class of $2^{100}$ ciphers.  SUC concept was presented recently as a physical security anchor to overcome the drawbacks of the traditional analog Physically Unclonable Functions (PUFs). Such unknown ciphers may be permanently self-created within System-on-Chip SoC non-volatile FPGA devices to serve as a digital clone-resistant structure. Moreover, a lightweight identification protocol is presented in open networks for physically identifying such SUC structures in FPGA-devices. The proposed new family may serve for lightweight realization of clone-resistant identities in future self-reconfiguring SoC non-volatile FPGAs. Such self-reconfiguring FPGAs are expected to be emerging in the near future smart VLSI systems.  The security analysis and hardware complexities of the resulting clone-resistant structures are evaluated and shown to exhibit scalable security levels even for post-quantum cryptography.
\end{abstract}

%INTRODUCTION
\section*{Introduction}

Clone-Resistant Units have been well investigated during the last two decades. The aim is to provide electronic systems with unique and secure identities making them resistant to cloning attacks. Physical Unclonable Functions (PUFs) \cite{Pappu2001}\cite{Adi2017}\cite{maes2010physically} were introduced to fabricate electronic unclonable units for secure identification/authentication \cite{tuyls2006rfid}\cite{Sadeghi2010}, memoryless key storage \cite{Lim2004a}\cite{vskoric2005robust} and intellectual property protection \cite{IntrinsicID}. Due to the analog nature of all proposed PUF technologies, all techniques proposed so far had limited use in real world applications due to economic cost factors and failing long term stability. To overcome the PUFs issues, digital clone-resistant functions were introduced in \cite{Adi2008a}. In \cite{Adi2017e}, digital clone-resistant functions were coined as Secret Unknown Ciphers. By definition, a Secret Unknown Cipher is a randomly, internally generated cipher inside the chip where the user has no access or influence on its creation process, even the producer is not able to backtrace the personalization process and deduce the made random cipher \cite{Mars2019a}\cite{Mars_V2X}. \\
Creating SUC requires designing families of secure ciphers with random components \cite{Mars2019a}. In \cite{rspnsuc}\cite{Mars2019a}, an SUC based on random block cipher was proposed, it is deploying random optimal S-Boxes as source of randomness of the SUC design. It was shown that deploying a fixed SUC design with small random components increases tremendously the cardinality of the resulting SUC class. Also, this ensures that each element of the SUC class has the same security level \cite{Mars2019a}. In \cite{Mars2017}, Random Stream Cipher (RSC) based on single cycle T-Functions  (Triangular Functions) has been proposed to construct a class of SUCs. The proposed RSC-SUC makes use of DSP blocks embedded in modern SoC FPGAs to implement single cycle T-Functions as part of the keystream generators. Both proposed designs in \cite{rspnsuc}\cite{Mars2019a} and \cite{Mars2017} share the property of requiring a small area that should be located inside the FPGA. Distributing the SUC design in both cases would results with more area overhead and latency. In this paper, the SUC design template is based on combining a set of NLFSRs. Distributing NLFSRs overall the FPGA area is practically attainable since each NLFSR can be implemented in a free area of the FPGA. This would ensure additionally a zero-cost implementation of the SUC and lower its vulnerability to some side channel attacks.\\
In 2005, the European project ECRYPT launches a competition to design new stream ciphers that might be suitable for widespread adoption. This project is called eSTREAM (ECRYPT Stream Cipher Project) \cite{ECRYPT} and it received 35 submissions. When it came to its end in 2008 \cite{Canniere2008}, four of the proposals in the final portfolio \cite{Babbage2008}
were suited to fast encryption in software: HC-128, Rabbit, Salsa20/12 and Sosemanuk, while other four stream ciphers offered particularly efficient hardware implementation: Grain v1, MICKEY 2.0, Trivium and F-FCSR-H which has been excluded later because of the cryptanalytic results presented in \cite{M.HellandT.Johansson2008}. The last eStream portfolio includes seven algorithms \cite{Technologies2009}. A number of NLFSR-based stream ciphers have been proposed to the eStream project such as Achterbahn \cite{Gammel2005} and Grain \cite{Hell2007}. Achterbahn was one of the challenging new designs based on combining several NLFSRs with a non-linear combining function, which performs nonlinear operations on sequences with distinct minimal polynomials \cite{Rueppel1986}. In \cite{T.Johansson2006}, authors  highlight some problems in the design principle of Achterbahn summarized in the small length of the NLFSRs and the weakness of the combining function. The complexity of the attack presented in \cite{T.Johansson2006} depends exponentially to the number of shift registers and their size, and to the number of shift registers outputs that would cancel the nonlinear part of the combining function if they are equal to zero. After selecting the positions in the output sequence that cancel the nonlinear terms in the combining function, the attack builds parity checks. We outline that a high number of shift registers in the linear part of the combining function makes the attack complexity very high.
\subsection*{Contribution.}
The contributions of this paper are summarized as follows:  firstly, a new large-class of low-complexity stream ciphers with the same designed-security level is created. Each resulting cipher, even when randomly selected, exhibits the same security level. The cardinality of the cipher-class exceeds $2^{100}$ without considering the NLFSRs initial states as a key of 223 bits. Secondly, the resulting ciphers are adapted to convert future VLSI-devices to clone-resistant physical entities in future VLSI technologies. Finally, a new generic-lightweight identification/authentication protocol is shown for VLSI-devices when using such SUC-based structure.\\
The remainder of this paper is organized as follows, section~\ref{sec:sec1} describes the state of the art of clone-resistan units, also it discusses Kerckhoffs' principles in relation to SUC. Section~\ref{sec:sec2} presents a detailed description of the key stream generator. In section~\ref{sec:sec3}, security analysis of the proposed family of new stream ciphers is investigated. Section~\ref{sec:sec4} describes a concept for deploying this family to create SUCs and provide unique and robust identity to SoC units. Section~\ref{sec:sec5} presents the hardware complexity results and Section~\ref{sec:sec6} concludes.

%1. Secret Unknown Cipher
\section{Clone-Resistant Units}
\label{sec:sec1}
\subsection{Physical Unclonable Functions}
Physical Unclonable Functions (PUFs) \cite{Adi2017}\cite{maes2010physically}\cite{Pappu2001} are increasingly proposed as central building block in cryptographic protocols and security architectures. They are proposed to be used for secure devices identification/authentication, memoryless key storage and intellectual property protection. Most PUFs responses are noisy and only contain a limited amount of entropy. Hence, they cannot be used as keys directly. To remedy this problem, fuzzy extractors \cite{Bosch2008} \cite{Dodis2004} \cite{vskoric2005robust} were proposed to be used beside each PUF. They are working on two steps: in the enrollment phase, a helper data is extracted by deploying a helper data algorithm. During the reconstruction phase, fuzzy extractor algorithm uses the helper data and the PUF response to reproduce the key. These error correction mechanisms are expensive and require high number of logic gates \cite{Bosch2008} \cite{Dodis2004}.\\
Furthermore, many attacks on PUFs have been proposed recently, they are targeting both weak PUFs and strong PUFs  \cite{guajardo2007fpga} \cite{ruhrmair2012security}; weak PUFs have few challenges, commonly only one challenge per PUF instance, hence it is assumed that the access to the weak PUF response is restricted. However, semi-invasive means have been used to reveal the state of memory-based PUF \cite{Nedospasov2013}. The second major PUFs types are Strong PUFs, they have large number of challenge-response pairs and they are unpredictable. Hence, protecting the challenge-response interface is not required. Strong PUFs are less susceptible to cloning and invasive attacks as weak PUFs. However, modeling attack constitutes a strong technique to clone strong PUFs, it has been introduced firstly by D. Lim to model an Arbiter-Based PUF \cite{Lim2004a} and later on by Matzoobi et al. to evaluate linear and feed-forward PUF structures \cite{Majzoobi2008}. Recently, Rührmair et al. demonstrate PUF modeling attacks on many PUFs by using machine learning techniques \cite{ruhrmair2013puf}\cite{ruhrmair2012security}\cite{ruhrmair2010modeling}. The attack succeeds if the adversary constructs an algorithm which behaves indistinguishably from the original PUF on almost all Challenge Response Pairs (CRPs). In \cite{Merli2011b}, side channel attack was used to analyze PUFs architecture and fuzzy extractor implementations by deploying power analysis. Recent trends combine both side channel and modeling attacks \cite{Delvaux2013} \cite{Mahmoud2013} to facilitate machine learning which is deployed in modeling attack.\\
\subsection{Secret Unknown Cipher}
\begin{center}
\begin{figure}
\includegraphics[scale=0.4]{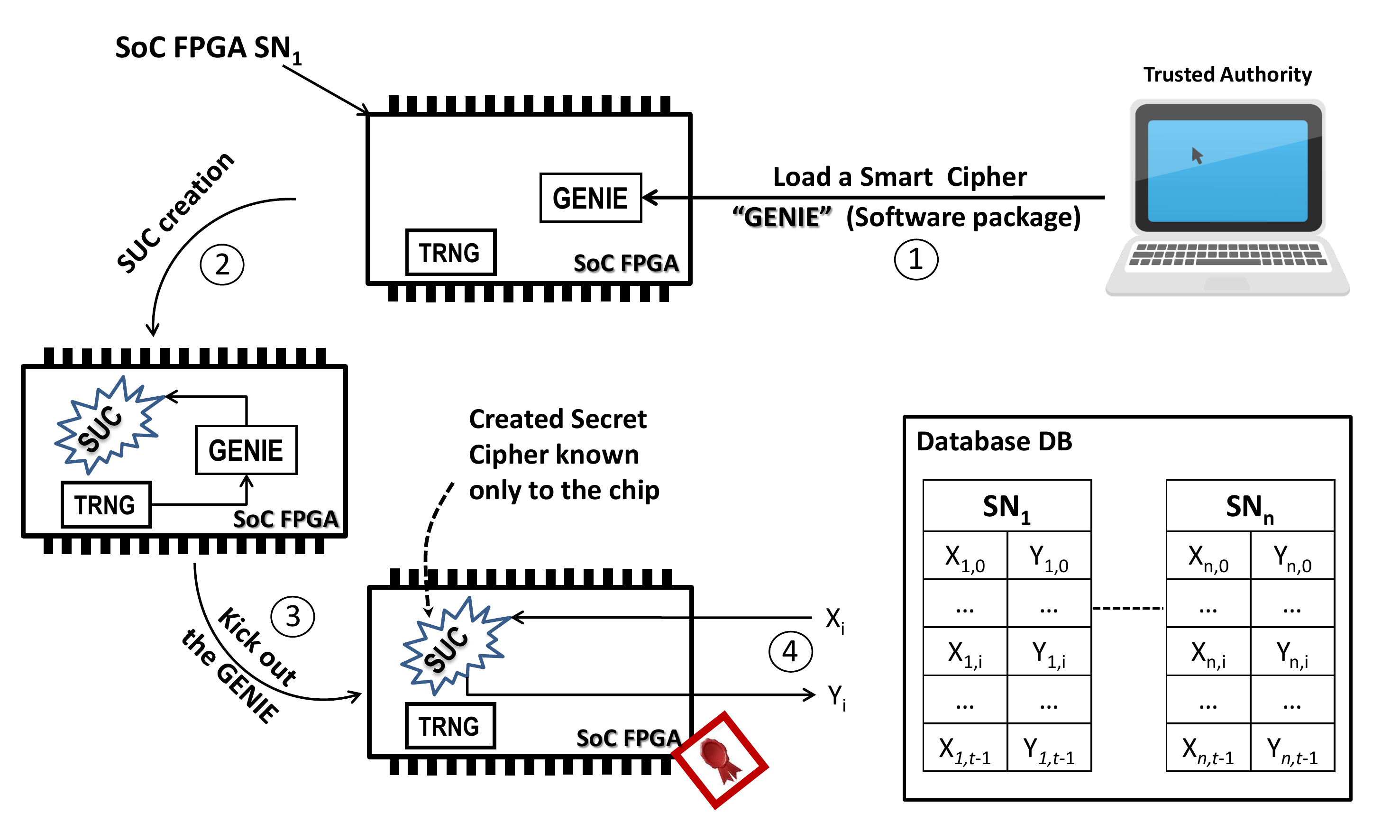}
\caption{Concept for creating SUC in SoC FPGAs environment}
\label{fig:SUC}
\end{figure}
\end{center}

Digital physical clone-resistant units based on pseudo-random functions have been proposed in \cite{Adi2007c}\cite{Adi2008a} to overcome some of the PUFs drawbacks especially their inconsistency. Those Physical Clone-Resistant Functions were nominated later on as Secret Unknown Ciphers (SUCs) \cite{rspnsuc}\cite{Mars2017}. 
\begin{definition}
Secret Unknown Cipher is a randomly internally generated cipher/hash inside the chip, where the user has no access or influence on its creation process, even the producer should not be able to back trace the creation process and deduce the made random cipher. Each generated SUC can be defined as an invertible Pseudo Random Function (PRF), as follows:
\begin{equation}\label{eq1}
\begin{split}
SUC: \{0,1\}^n &  \longrightarrow \{0,1\}^m   \\
X & \xrightarrow{PRF}Y
\end{split}
\end{equation}
and 
\begin{equation} \label{eq2}
\begin{split}
SUC^{-1}: \{0,1\}^m & \rightarrow \{0,1\}^n  \\
Y & \xrightarrow{PRF^{-1}}X
\end{split}
\end{equation}
For an SUC based on block cipher design, i.e. $n=m$. The optimum case is to design an involutive SUC, such as  $SUC=SUC^{-1}$ and hence we define it as follows:

\begin{equation}\label{eq3}
\begin{split}
 & SUC: \{0,1\}^n \rightarrow \{0,1\}^n \\
        & \text{where }  SUC(SUC(X))=X  \text{ for   all }   X\in\{0,1\}^n
\end{split}
\end{equation}

\end{definition}
Figure~\ref{fig:SUC} describes the concept for embedding SUC in System on Chip (SoC) FPGA devices. The personalization process proceeds as follows, the Trusted Authority (TA) disposes of a software package called “GENIE” that contains an algorithm for creating internally random secure ciphers in addition to a package of cryptographically strong functions that will be used to design each SUC. The TA injects for a short time into each SoC FPGA unit the GENIE which runs only one time (step 1). After being loaded into the chip, the GENIE creates a permanent (non-volatile) and unpredictable random cipher by deploying random bits from the True Random Number Generator (TRNG) (step 2).  When the GENIE completes the creation of the SUC, it will be fully deleted (step 3). After that, the SoC FPGA will contain its unique and unpredictable SUC. The TA challenges the SUC by a set of challenges $X_i$ and gets the corresponding responses $Y_i=SUC(X_i)$ and stores them on the corresponding area in its Units Individual Records (UIR) defined by the Serial Number of the device $SN_i$. 
The X/Y pairs are to be used later by a TA to identify and authenticate devices. The concept is comparable to a PUF with the advantage that with an SUC based on random block cipher design, recovering Y from X is possible by using the inverse function of the SUC ($SUC^{-1}$) or the same involution SUC. This property was deployed in \cite{Mars_V2X} to build a chain of trust used for a secured vehicular over the air software update, also for securing in-vehicle and vehicle-to-vehicle communication. SUC invertability is also used in the generic identification and authentications protocols in \cite{Adi2017e} and other applications protocols such as in \cite{newecoin}.
In \cite{Adi2017e}, two generic identification/authentication protocols have been proposed, they show a very efficient identification protocol which eliminates storage of big number of challenge/response pairs, also the device should embed a small memory of about $t$-bit to detect $t$-consumed $X/Y$ pairs, even without the need of deleting the used pairs during communication with the unit.\\
In \cite{rspnsuc}\cite{Mars2019a}, template based SUC was presented, where a block cipher with random components was designed as an SUC template. Optimal 4-bit S-Boxes were used as a source of randomness, such as the GENIE selects few S-Boxes from some sets of all optimal 4-bit S-Boxes. Each resulting SUC from this class has the same security level. Furthermore, in \cite{Mars2017}, Mars et al. proposed the first digital clone-resistant function prototype based on Random Stream Cipher (RSC) deploying a class of T-Functions (Triangular Functions) as key stream generator. We note that, identification and authentication protocols designed for SUC based on random block ciphers would not be applicable directly for an SUC based on random one-way function or on RSC.\\

As discussed before, many PUFs are susceptible to mathematical cloning hence they are recently nominated as Physically Unclonable Functions. SUC designs ensure that it is secure against known mathematical cryptanalysis as in \cite{Mars2019a}\cite{Mars2017}, and each SUC have the same security level. Since each device embeds a unique SUC, the adversary should break each unit alone with the same attack complexity. Moreover, SUC can be implemented with zero cost; the hardware overhead of an SUC should be low such as in \cite{Mars2019a}. Most industrial customer designs do not make full usage of the FPGA resources. Hence with low overhead, SUC can make use of the free FPGA resources and can be incrementally added to the customer design with zero cost. \\

\subsection{Kerckhoffs' principles and SUC}
In \cite{Kerckhoffs1883}, Kerckhoffs stated the principles that should apply to a cryptosystem. The most concerning one, in relation to SUC, is the one stating that the method used to encipher data is known to the opponent, and that security must lie in the choice of key. However, \textit{''This does not necessarily implies that the method should be public, but only considered as public during its creation''} \cite{Kerckhoffsonline}\cite{petitcolas1883cryptographie}. Thus, SUC validates this Kerckhoffs's principle if and only if the SUC design is secure when considering that all the components are publicly known. In this paper, a family of stream cipher is proposed such that the NLFSRs feedback functions are selected randomly together with the initial NLFSRs states to generate SUCs. The security analysis of the proposed family of stream ciphers is investigated by considering that the cipher design is publicly known. i.e. the NLFSRs's feedback functions are known. \\

\noindent Cryptanalyzing SUCs in the field would require two steps:
\begin {itemize}
\item \textit{Reversing the secret components}: an adversary is forced to reverse the random selected functions that are  used by the SUC.
\item \textit{Breaking the resulting stream cipher}: After reversing the secret parameters of an SUC, this SUC could be considered as a publicly known cipher and an adversary would apply known cryptanalytical attack to break this SUC. 
\end {itemize}

Since each SUC is assumed to be generated randomly, attacking each SUC requires to repeat the same attack with the same complexity. This constitutes an advantage over using any secure stream cipher with publicly known specifications and with a randomly generated secret key. For this latest solution, the attack complexity is only based on breaking the publicly known stream cipher. Another advantage by design, is that for the proposed SUC, the secrets are distributed and not located in the same area such that when deploying a random secret key, this makes it hard to physically break the SUC.

%2. Description of the Keystream Generato
\section{Description of the Keystream Generator}
\label{sec:sec2}
The basic components of the KSG are 16 Non-Linear Feedback Shift Registers (NLFSRs) of lengths 6 to 17 and 19, 21, 22 and 23, combined by a balanced Boolean function F with algebraic degree 4, correlation immunity 8 and algebraic immunity 4. The NLFSRs are such that they can produce binary sequences of maximum period $2^N - 1$, where $N$ is the length of the shift register. Each shift register has a corresponding set of non-linear feedback function updating the internal state of the shift register. The outputs of the 16 NLFSRs deliver the 16 inputs of the combining function $F$ which outputs the running key $Z_t$. The 5 4-bits Look Up Tables (4-LUTs) implement the Boolean combining function $F$. The total number of all the NLFSRs bits is 223 bits. This design is hardware oriented to FPGA environment where the basic logic computing unit in FPGA is 4-LUT.  \\
The key-loading algorithm that determines the initial internal states of the NLFSRs from a key ($K \geq 80$ bits) and an initial vector (IV) will not be investigated in this paper since it is not required for SUC usage, i.e. the key (K) can be the initial NLFSRs states. Generally, two requirements are important for the key loading algorithm: for all possible keys (K), the key loading algorithm should generate non-zero initial states for the NLSRs, also it should be resistant to side channel attacks.
%The proposed KSG belongs to the category of synchronous stream ciphers.

\begin{figure}[H]
\begin{center}
\includegraphics[scale=0.5]{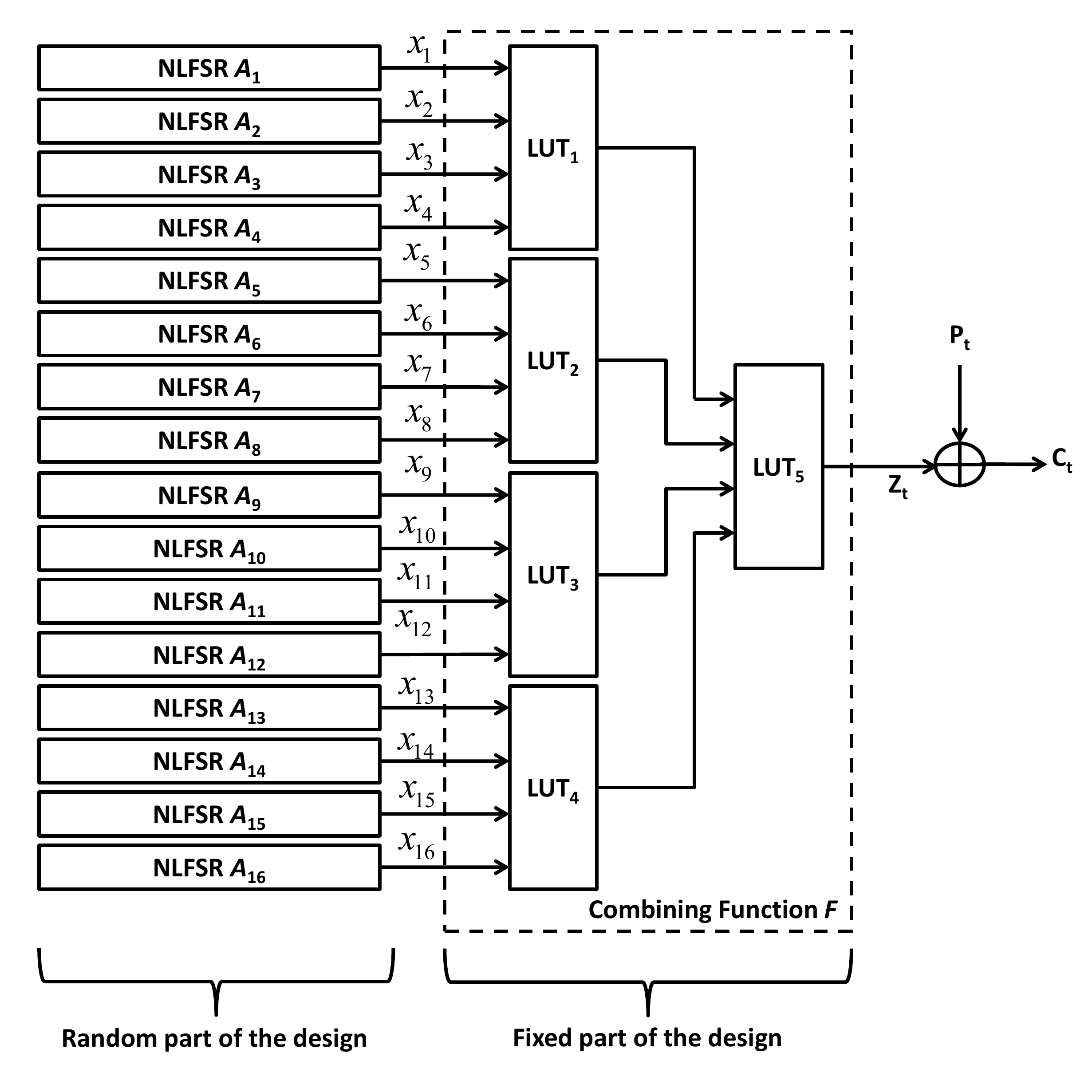}
\caption{Description of the keystream generator}
\label{fig:KSG}
\end{center}
\end{figure}

\noindent Figure~\ref{fig:KSG} describes the proposed stream cipher design.

%2.2_NLFSRs
\subsection{Non-Linear Feedback Shift Registers}
The principal components of KSG are the 16 NLFSRs with lengths from 6 to 17 and 19, 21, 23 and 24. Each $N$-bit NLFSR has a set of feedback functions ensuring all a maximum period of $2^N-1$. This section will describe in details the NLFSRs design methodology.\\

\begin{definition}
A feedback Shift Register consists of pure cycles if and only if its feedback function has the form:
\begin{equation}
\label{eq:basicform}
f(x_0,x_1,...,x_{N-1})=x_0\oplus g(x_1,...,x_{N-1})
\end{equation}
where g is a Boolean function that does not depend on $x_0$.
\end{definition}

\definition{A (binary) \textit{de Bruijn sequence} is a sequence of period $2^N$ in which each $N$-bit pattern occurs exactly once in one period of the sequence.\\
\noindent The linear complexities of order $N$ de Bruijn sequences are bounded by $2^{N-1}+N$ and $2^N-1$ \cite{chan1982complexities}.}

\noindent There are $2^{2^{N-1}-N}$ different $N$-bit Fibonacci NLFSRs with the period $2^N$ \cite{fredricksen1982survey}.

\definition{A modified de Bruijn sequence is a sequence of period $2^N-1$ in which each $N$-bit pattern occurs exactly once in one period of the sequence.}\\

\par In \cite{Dubrovaa}, a list of maximum period  ($2^{N}-1$) NLFSRs of $N$-bits with $4 \leq N \leq24$  has been presented. The search covers three types of feedback functions with algebraic degree two:

\begin{itemize}
  \item $f_1(x_0,...,x_{N-1})=x_0\oplus g_1(x_a,x_b,x_c,x_d)=x_0\oplus x_a \oplus x_b \oplus x_cx_d$
  \item $f_2(x_0,...,x_{N-1})=x_0\oplus g_2(x_a,x_b,x_c,x_d,x_e)=x_0 \oplus x_a \oplus x_bx_c \oplus x_dx_e$
  \item $f_3(x_0,...,x_{N-1})=x_0\oplus g_3(x_a,x_b,x_c,x_d,x_e,x_h)=x_0 \oplus x_a \oplus x_b \oplus x_c \oplus x_d \oplus x_ex_h$
\end{itemize}
Where $a,b,c,d,e,h\in{1,2,...,N-1}$.\\
\noindent The presented NLFSRs in \cite{Dubrovaa} do not include the all-0 state in their longest cycle of states of period $2^N-1$.\\

The set of $N$-bit Fibonacci NLFSRs with the period $2^N-1$ can be partitioned into 4 subsets \cite{jansen1989investigations}: basic, reverse of basic, complement of basic, and reverse complement of basic.
In \cite{Dubrovaa}, only NLFSRs with basic form were listed. The forms of the reverse, complement and reverse complement of the basic form (equation~\ref{eq:basicform}) are described as follows:

\begin{itemize}
\item Reverse form: $f_r(x_0,x_1,...,x_{N-1})=x_0\oplus g(x_{N-1},...,x_1)$
\item Complement form: $f_c(x_0,x_1,...,x_{N-1})=x_0\oplus 1 \oplus g(x_1,...,x_{N-1})$
\item Reverse complement form: $f_{rc}(x_0,x_1,...,x_{N-1})=x_0 \oplus 1 \oplus g(x_{N-1},...,x_1)$
\end{itemize}
Thus, for each listed feedback function in \cite{Dubrovaa}, three feedback functions generating the reverse, complement or reverse complement sequence can be deduced.\\

In \cite{Dubrovaa}, for each NLFSR $A_i$ with $N_i$-bit where $4 \leq N_i \leq 24$, a set of feedback functions ensuring maximum period of $2^{N_i}-1$ was presented. All feedback functions have the form in equation~\ref{eq:basicform}. For NLFSRs with $N_i$-bit, $S_{N_i}$ denotes the set of boolean functions $g$ listed in \cite{Dubrovaa} (by removing the xor with $x_0$) together with their reverse, complement and reverse complement form. We coin those functions as Random Feedback Functions ($RFF_{N_i}$). The set $S_{N_i}$ with only the functions having basic form (without xor with $x_0$) are listed in Table~\ref{table:nlfsrs}.

%The feedback functions $f$, $f_r$, $f_c$ and $f_{rc}$ have all Boolean functions $G$ that their output is xored with $x_0$ to generate the feedback bit.

%For an NLFSR $A_i$ with $N_i$-bit, a set $S_{A_i}$ of Boolean functions $G$ ensuring that $A_i$ has maximum period $2^{N_i}-1$ is selected. We coin those functions $G$ as Random Feedback Functions ($RFF_{N_i}$). The set $S_{A_i}$ with only the functions having basic form and ensuring a maximum period of $2^{N_i}-1$ will be listed in Table~\ref{table:nlfsrs}.\\

%{\color{red}In \cite{almuhammadi2018nlfsr}, authors investigated the same type of feedback functions as in \cite{Dubrovaa} and resulted with additional feedback functions. The results from \cite{Dubrovaa} and \cite{almuhammadi2018nlfsr} are combined in Table~\ref{table:nlfsrs} for only the used NLFSRs lengths.}

\begin{figure}%[b]
\begin{center}
\includegraphics[scale=0.5]{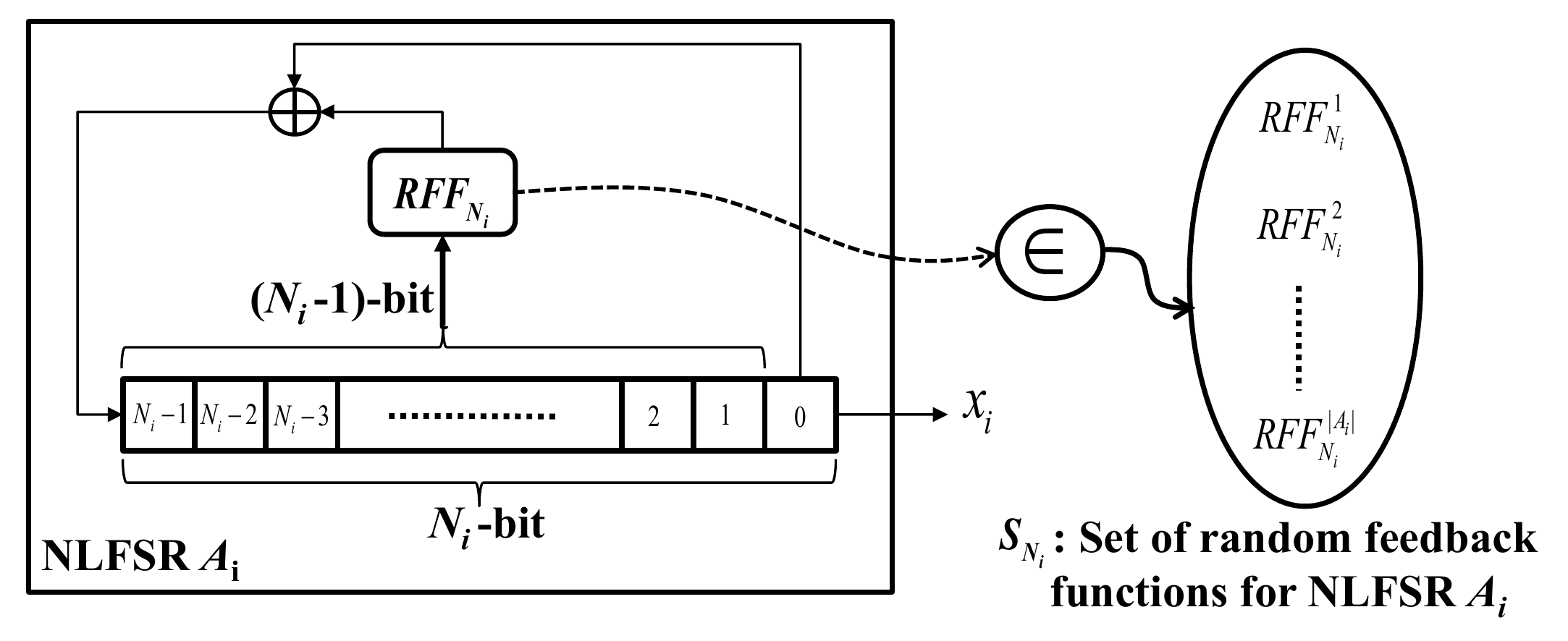}
\caption{General structure of the NLFSRs}
\label{fig:one_nlfsr}
\end{center}
\end{figure}

Figure~\ref{fig:one_nlfsr} describes the general structure of the used NLFSRs. For each NLFSR $A_i$ of length $N_i$, the feedback function contains Random Feedback Function ($RFF_{N_i}$). Its general form is defined as follows:
\begin{equation}
f(x_0,x_1,...,x_{N_i-1})=x_0 \oplus RFF_{N_i}(x_1,...,x_{N_i-1})
\end{equation}
Where:
\begin{itemize}
\item For each NLFSR $A_i$ of length $N_i$, a set of random feedback functions $S_{N_i}$ is selected such that each of its $RFF_{N_i}^{j}$ makes the NLFSR $A_i$ achieves a maximum period of $2^{N_i}-1$. where:

\begin{equation}
RFF_{N_i}\in S_{N_i}=\{RFF_{N_i}^{1}, ..., RFF_{N_i}^{j}, ..., RFF_{N_i}^{|A_i|}\}
\end{equation}

During the personalization process, one of the feedback functions is to be selected randomly from this set and will be used to construct NLFSR $A_i$. 

%The set $S_{N_i}$ with only the functions having basic form and ensuring a maximum period of $2^N-1$ will be listed in Table~\ref{table:nlfsrs}.\\

\end{itemize}

Each selected NLFSR $A_i$ in Figure~\ref{fig:KSG} has a form of the general structure in Figure~\ref{fig:one_nlfsr}, and generates a nonlinear sequence of period $2^{N_i}-1$ which is a nonlinear \textit{modified de Bruijn sequence}. The linear complexity $L_i$ of an NLFSR $A_i$ is bounded by: 

\begin{equation}
2^{N_i-1}+N_i \leq L_i \leq 2^{N_i}-1
\end{equation}

The number of NLFSRs and their lengths are selected to satisfy the primarily security requirements described in the following: 
\begin{itemize}
\item B-M Algorithm attack: in order to ensure that the attack complexity of B-M Algorithm is over $2^{80}$; in terms of time complexity, the linear complexity $L$ of the total key stream sequence should exceed $2^{40}$, since the complexity of B-M algorithm attack is the square of the linear complexity. For the data complexity, it is $2L$ and hence $L$ should be greater than or equal to $2^{80}$.
\item Correlation immunity: If an adversary succeeds to recover the randomly selected feedback functions, she/he can try to mount a correlation attack. In this case, to protect the system against correlation attack, the total number of bits of certain number of the least NLFSRs, where the number is determined by the correlation immunity, should exceed the computing power namely $2^{80}$. The correlation immunity of the combining function $F$ is equal to 8, thus the total number of bits of the shortest 9 NLFSRs should exceed 80. 
\end{itemize}

\noindent With these two constraints, we obtain the optimal NLFSRs designs for the proposal.    \\

Table~\ref{table:nlfsrs} describes the sets $S_{N_i}$ of the selected random feedback function $RFF_{N_i}$ that have only the basic form, the reverse, complement and reverse complement forms can be deduced easily. In the format of the $RFF_{N_i}$, indexes of variables of each product-term of a feedback function are separated by a comma, round bracket around the indexes denotes that those indexes belong to the same product-term. For example, 1,2,(2,4) represent the $RFF_{N_i}$:
\begin{equation}
RFF_{N_i}(x_1,x_2,x_3,x_4,x_5)=x_1+x_2+x_2x_4
\end{equation}

%\newpage
\begin{center}
\begin{longtable}{|c|c|c|}
\caption{Sets of selected random feedback functions for each NLFSR length}
\label{table:nlfsrs}\\
\hline
\textbf NLFSR & Length $N_i$  & Set of Random Feedback Functions $S_{N_i}$ \\[0.5ex]
\hline
\endfirsthead
\multicolumn{3}{c}%
{\tablename\ \thetable\ -- \textit{Continued from previous page}} \\
\hline
\textbf NLFSR & Length $N_i$ & Set of "basic" Random Feedback Functions $S_{N_i}$ \\[0.5ex]
\hline
\endhead
\hline \multicolumn{3}{r}{\textit{Continued on next page}} \\
\endfoot
\hline
\endlastfoot

%\hline
 $A_1$ & $6$ & 
%6
\thead{
1,2,(1,2); 1,2,(2,4); 1,3,(1,5); 1,4,(1,4); 2,3,(1,3);2,3,(1,5); 2,3,(2,3);\\
2,3,(2,4); 1,(1,2),(4,5); 1,(1,3),(3,5); 1,(2,3),(2,5); 2,(1,3),(2,4);\\
2,(1,3),(3,4); 2,(1,3),(3,5); 2,(1,5),(2,4); 2,(1,5),(4,5); 2,(2,3),(3,5);\\
2,(3,4),(3,5); 3,(1,4),(2,3); 3,(1,4),(2,4); 3,(1,4),(3,4);
} \\
 \hline
 $A_2$ & $7$ &    
%7
\thead{
1,2,(2,6); 1,4,(1,3); 1,5,(1,5); 1,5,(3,5); 1,5,(4,6); 2,4,(1,2); 2,4,(2,5);\\
1,(1,2),(5,6); 1,(1,5),(3,4); 1,(1,6),(4,5); 1,(2,3),(3,5); 1,(2,5),(3,5);\\ 
1,(2,5),(4,5); 1,(3,4),(4,5); 2,(1,2),(4,6); 2,(1,4),(3,4); 2,(1,5),(2,6);\\
2,(1,6),(2,4); 2,(1,6),(3,6); 2,(1,6),(5,6); 2,(2,4),(3,5); 2,(2,5),(4,6);\\
2,(2,6),(4,6); 2,(3,6),(5,6); 3,(1,2),(2,3); 3,(1,3),(1,6); 3,(1,4),(3,6);\\
3,(1,5),(3,5); 3,(1,6),(3,4); 3,(2,3),(4,5); 3,(2,5),(3,5); 1,2,3,4,(1,6);\\
1,2,3,4,(2,3); 1,2,3,4,(2,6); 1,2,3,6,(1,3); 1,2,3,6,(1,5); 1,2,3,6,(2,6);\\
1,2,4,5,(1,2); 1,2,4,5,(1,5); 1,2,4,5,(2,6)
}\\

\hline
 $A_3$ & $8$ &   

%8
\thead{
1,5,(1,5); 1,6,(1,2); 1,6,(1,7); 1,6,(2,4); 1,6,(4,5); 1,6,(5,6); 2,5,(2,4);\\
2,5,(3,7); 2,5,(4,5); 3,4,(2,4); 3,4,(2,7); 3,4,(3,4); 3,4,(4,6); 3,4,(4,7); \\
3,4,(6,7); 1,(1,4),(2,4); 1,(1,6),(2,5);1,(2,3),(2,4); 1,(2,4),(6,7); \\
1,(3,4),(4,7); 2,(1,3),(4,6); 2,(1,3),(5,7); 2,(1,5),(6,7); 2,(1,7),(2,3); \\
2,(3,7),(6,7); 3,(1,2),(2,4); 3,(1,4),(2,4); 3,(1,6),(3,6); 3,(1,6),(4,6); \\
3,(1,6),(4,7); 3,(2,3),(5,6); 3,(2,4),(6,7); 3,(2,6),(3,7); 1,2,3,5,(2,6); \\
1,2,3,6,(3,5); 1,2,3,6,(5,7); 1,2,4,5,(2,4); 1,2,4,7,(1,5); 1,2,5,7,(2,4); \\
1,3,4,7,(1,4); 1,3,4,7,(1,6); 1,3,4,7,(3,7)
}\\

 \hline
 $A_4$ & $9$ &   

%9
\thead{
1,6,(4,6); 1,6,(4,8); 2,4,(4,5); 3,4,(3,7); 1,(1,5),(2,5); 1,(1,6),(6,7);\\
1,(1,8),(2,7); 1,(1,8),(5,6); 1,(2,3),(3,8); 1,(2,8),(3,7); 1,(3,4),(3,5);\\
1,(3,7),(5,8); 2,(1,5),(4,6); 2,(1,6),(2,7); 2,(1,8),(3,4); 2,(2,7),(4,6);\\
2,(4,7),(5,6); 3,(1,2),(4,7); 3,(1,6),(1,7); 3,(1,7),(4,8); 3,(2,3),(4,7);\\
4,(1,3),(2,8); 4,(1,6),(3,6); 4,(2,3),(5,8); 4,(2,5),(2,8); 4,(2,7),(3,8);\\
4,(2,8),(6,7); 4,(3,5),(3,7); 1,2,3,4,(3,7); 1,2,3,7,(4,6); 1,2,4,7,(1,6);\\
1,2,5,6,(1,6); 1,2,5,6,(2,6); 1,2,5,8,(2,6); 1,2,6,7,(3,6); 1,3,4,5,(3,7);\\
1,3,5,7,(5,6); 1,3,5,8,(3,5); 1,4,6,7,(1,7); 2,3,4,7,(2,8)
}\\
\hline

$A_5$ & $10$ &    

%10
\thead{
1,2,(8,9); 1,4,(3,7); 1,8,(6,7); 2,5,(1,5); 4,5,(2,6); 4,5,(4,8); 4,5,(4,9);\\
1,(1,2),(3,4); 1,(2,4),(2,5); 1,(2,8),(7,9); 1,(3,8),(4,7); 1,(4,8),(6,7);\\
2,(1,3),(4,7); 2,(1,4),(3,7); 2,(1,5),(3,5); 2,(1,5),(4,9); 2,(1,6),(1,7);\\ 
2,(1,7),(4,6); 2,(1,9),(5,9); 2,(3,5),(3,7); 2,(3,9),(8,9); 3,(1,2),(2,8);\\
3,(1,3),(7,9); 3,(1,6),(3,8); 3,(1,6),(6,9); 3,(2,3),(2,6); 3,(2,7),(8,9);\\
3,(2,8),(7,9); 3,(6,7),(8,9); 4,(1,3),(1,7); 4,(1,3),(7,8); 4,(1,3),(7,9);\\
4,(1,5),(1,9); 4,(1,5),(7,9); 4,(7,8),(7,9); 1,2,4,8,(1,5); 1,2,4,8,(2,4);\\  
1,2,5,8,(5,9); 1,3,4,7,(3,6); 1,3,6,7,(1,6); 1,4,5,9,(1,9); 1,4,5,9,(4,9);\\
1,4,5,9,(5,9); 1,5,6,7,(2,8); 2,3,4,6,(3,6); 2,4,5,8,(2,4); 2,4,6,7,(1,6)
}\\
 \hline
 $A_6$ & $11$ &   

%11
\thead{
1,9,(1,4); 2,5,(1,9); 2,8,(6,9); 1,(1,7),(2,8); 1,(1,9),(2,7); 1,(2,3),(4,5);\\
1,(2,5),(3,4); 1,(2,7),(3,10); 1,(3,7),(3,8); 1,(3,7),(7,8); 2,(4,5),(6,10);\\
2,(4,6),(9,10); 2,(7,9),(8,10); 3,(1,6),(8,9); 3,(1,9),(5,10); 3,(2,7),(5,7);\\ 
3,(3,5),(6,9); 3,(3,6),(5,8); 3,(3,7),(7,10); 4,(1,2),(9,10); 4,(2,3),(2,10);\\
4,(3,7),(4,8); 5,(1,4),(6,9); 5,(2,8),(6,8); 5,(4,7),(6,7); 1,2,3,5,(4,6); \\
1,2,4,5,(4,6); 1,2,4,7,(2,3); 1,2,4,7,(4,9); 1,2,4,7,(8,9); 1,2,4,10,(1,9); \\
1,2,4,10,(3,9); 1,2,7,8,(1,9); 1,2,7,8,(9,10); 1,3,4,10,(6,10); 1,3,6,8,(6,8);\\
1,3,6,10,(7,9); 1,3,7,9,(1,8); 1,4,5,8,(5,7); 1,4,7,10,(1,9); 1,5,6,8,(5,9);\\
1,5,7,9,(2,8); 1,6,8,9,(2,6); 2,3,7,8,(4,10); 2,3,7,8,(6,10); 2,3,7,8,(7,10);\\
2,4,5,9,(5,9); 3,4,5,6,(2,10); 3,4,6,7,(2,3); 3,5,6,7,(4,8)
}\\

\hline

$A_7$ & $12$ &    

%12
\thead{
3,8,(3,9); 4,7,(1,7); 4,7,(4,7); 1,(2,3),(3,4); 1,(2,5),(3,10); 1,(2,8),(6,10);\\
1,(7,8),(8,10); 1,(8,11),(9,10); 2,(1,3),(3,6); 2,(1,7),(2,8); 2,(1,10),(1,11);\\
2,(2,3),(7,9); 2,(3,9),(3,11); 2,(3,9),(5,9); 2,(5,11),(8,11); 2,(7,9),(7,11); \\
3,(1,8),(7,10); 3,(5,11),(6,10); 1,2,3,5,(5,9); 1,2,5,9,(7,11); 1,2,6,11,(2,6);\\
1,3,6,7,(4,10); 1,3,6,9,(1,9); 1,3,6,9,(4,10); 1,3,7,10,(4,5); 1,4,8,10,(2,5);\\ 
1,5,6,8,(4,6); 1,5,6,8,(6,10); 1,5,6,11,(7,8); 1,5,7,9,(1,11); 1,5,9,10,(6,7);\\
2,3,4,10,(3,8); 2,3,6,8,(3,6); 2,3,6,10,(2,6); 2,3,6,10,(4,10); 2,5,6,10,(2,10)
}\\

\hline
$A_8$ & $13$ &    

%13
\thead{
1,11,(5,9); 4,8,(9,10); 1,(1,7),(3,7); 1,(2,3),(6,11); 1,(2,5),(5,11);\\
1,(2,6),(6,8); 1,(2,9),(4,5); 2,(1,6),(9,12); 2,(7,10),(10,12); 3,(1,9),(2,11);\\
3,(4,6),(9,11); 3,(8,9),(9,10); 4,(1,3),(4,6); 4,(1,3),(10,12); 4,(2,9),(8,10);\\
5,(1,5),(4,9); 5,(1,12),(7,11); 5,(2,9),(4,5); 5,(3,6),(4,9); 5,(3,12),(9,11);\\
6,(1,5),(2,12); 1,2,4,5,(1,7); 1,2,10,11,(6,12); 1,3,4,6,(6,10); 1,4,5,10,(4,8);\\
1,5,6,7,(5,9); 1,5,7,9,(8,9); 1,5,7,11,(8,10); 1,7,10,11,(2,6); 1,8,9,10,(8,9);\\
2,3,8,11,(1,10); 2,5,6,11,(8,11); 2,6,7,10,(8,12); 3,4,5,12,(4,5);\\
3,5,6,10,(8,11); 3,5,7,10,(2,10)
}\\

 \hline
 $A_9$ & $14$ &    
%14
\thead{
1,2,(7,12); 1,(2,13),(4,12); 1,(5,12),(9,12); 2,(1,5),(3,11);\\
3,(1,6),(4,12); 3,(2,4),(6,12); 3,(2,12),(6,13); 3,(5,10),(7,12);\\
5,(2,4),(6,13); 6,(1,13),(5,9); 6,(5,9),(12,13); 1,2,3,5,(1,3);\\
1,2,4,7,(1,3); 1,4,5,8,(2,8); 1,4,5,13,(1,6); 1,4,7,11,(1,11);\\
1,6,10,12,(3,7); 1,6,10,12,(7,9); 1,7,9,12,(3,13); 2,3,5,7,(1,5);\\
2,3,10,12,(9,10); 2,5,6,12,(6,10); 2,7,9,11,(11,12);\\
4,5,6,8,(1,4); 4,6,7,10,(5,13)
}\\
%\newpage
 \hline
 $A_{10}$ & $15$ &    

%15
\thead{
5,9,(2,11); 2,(6,8),(12,14); 4,(2,11),(7,10); 4,(5,6),(5,14);\\
4,(6,10),(9,10); 4,(7,8),(12,14); 6,(8,11),(12,13); 7,(2,11),(10,13);\\ 
7,(3,12),(3,13); 1,3,7,11,(9,10); 1,4,5,12,(3,4); 1,4,6,11,(2,14);\\
1,4,9,10,(7,10); 1,5,11,13,(5,11); 2,3,9,10,(6,10); 2,3,9,13,(3,7);\\
2,4,10,14,(4,10); 3,4,5,10,(3,7); 3,5,7,8,(3,13); 4,5,7,10,(1,14);\\
4,8,12,14,(5,6); 4,9,11,14,(1,13); 5,6,11,14,(5,8); 5,6,12,13,(5,9)
}\\

\hline
$A_{11}$ & $16$ &   
 
%16
\thead{
2,13,(2,3); 3,(1,5),(5,7; 3,(2,13),(7,14); 5,(4,8),(6,12); 5,(4,12),(7,8); \\
7,(2,6),(10,13); 7,(8,14),(11,12); 1,2,3,9,(6,14); 1,5,13,14,(14,15);\\ 
1,11,12,13,(5,15); 2,5,10,14,(6,14); 2,6,11,12,(14,15); 2,7,8,10,(3,6);\\
2,7,8,13,(3,15); 4,8,9,10,(8,12)
} \\

 \hline
 $A_{12}$ & $17$ &   

%17
\thead{
1,(7,10),(9,15); 3,(6,9),(13,14); 5,(4,7),(6,13); 6,(2,9),(7,12); 7,(1,8),(9,14);\\
8,(10,12),(11,16); 1,3,9,12,(7,13); 1,3,12,14,(2,10); 1,5,9,11,(1,13); \\
1,7,11,13,(6,14); 2,4,9,12,(6,16); 3,6,7,10,(9,15); 3,8,11,12,(3,11); \\
4,6,10,16,(3,11); 5,6,9,14,(6,14)
}\\
%\newpage
 \hline
 $A_{13}$ & $19$ &    

%19
\thead{
7,10,(6,18); 9,12,(1,13); 2,(6,8),(8,10); 4,(5,16),(7,14); 6,(4,8),(17,18);\\
1,4,5,8,(5,15); 1,4,8,17,(1,13); 3,7,9,16,(3,17); 5,6,12,14,(2,18)
}\\

 \hline
 $A_{14}$ & $21$ &    
 
 %21
\thead{
1,15,17,19,(13,15); 2,7,12,17,(4,10); 3,5,9,13,(15,17); 4,8,9,11,(3,11)
}\\

 \hline
 $A_{15}$ & $22$ &    

 %23
\thead{
1,(4,10),(11,18); 5,(4,12),(7,14); 1,6,8,12,(10,17);\\
1,10,16,18,(3,21); 5,6,11,15,(9,21)
}\\

\hline
 $A_{16}$ & $23$ &    

 %23
\thead{
3,(13,19),(18,19); 2,6,10,14,(5,13); 3,11,16,18,(4,19)
}\\
  
%\hline

\end{longtable}

\end{center}

\clearpage

  %%changes made here

%\textbf{The selected feedback functions can be implemented efficiently in FPGA environment by deploying 4-bit LUTs (4-LUT). Feedback functions with 4 variables, such as 0,1,2,(2,4), are implemented within one 4-LUT and the ones with more than 4 variables, such as 0,1,6,8,12,(10,17), are implemented with two 4-LUTs each. Hence, the maximum resources usage of the feedback functions is 32 4-LUTs.}

%%%CARDINALITY
\subsection{Cardinality of the KSG}
The proposed KSGs can be used to create a family of SUC. In general, this design randomness is based on deploying all possible feedback functions ensuring that the $N$-bit NLFSR generates a sequence with period $2^N-1$.
\noindent There are $2^{2^{N-1}-N+1}$ different $N$-bit Fibonacci NLFSRs with the period $2^N-1$ \cite{fredricksen1982survey}. Hence, the following theorem can be deduced.

\begin{theorem}
Let $N_i$ be the lengths of the NLFSRs of the KSG, where  $N_i\in S$ such as $S=\{6,...,17,19,21,22,23\}$ in this case.\\
The cardinality of the KSG deploying all Fibonacci $N_i$-NLFSRs with period $2^{N_i}$is:

\begin{equation}
\varsigma=2^{\sum_{N_i\in S}2^{N_i-1}-N_i+1}
\end{equation}
\end{theorem}

In this paper, the NLFSRs selected for the proposed design can be made random, since for each $N$-bit NLFSR there exist a number of possible feedback functions ensuring a maximum period of $2^N$. Hence randomly selecting one of the feedback functions for each NLFSR $A_i$ will ensure the same security level of the resulting random KSG as will be investigated through this paper.

\noindent Table~\ref{table:cardinality} presents the number of NLFSRs $|A_i|$ for each used NLFSR $A_i$ with length $N_i$:

\begin{table}[H]
\begin{center}
\caption{Number of all selected NLFSRs}
\label{table:cardinality}
\begin{tabular}{|l|l|l|l|l|l|l|l|l|}
\hline
$i$ & 1   & 2   & 3   & 4   & 5  & 6  & 7  & 8 \\ \hline
$|A_i|$ & 84  & 160 & 168 & 160 & 188 & 200 & 144 & 144 \\ \hline
\hline
$i$     & 9   & 10  & 11  & 12  & 13 & 14 & 15 & 16 \\ \hline
$|A_i|$ & 100 & 96  & 60  & 60  & 36 & 16 & 20 & 12  \\ \hline
\end{tabular}
\end{center}
\end{table}

\begin{theorem}
Let $N_i$ be the lengths of the NLFSRs of the KSG, where  $N_i\in S$ such as $S=\{6,...,17,19,21,22,23\}$.\\
We denote by  $|A_i|$ the number of NLFSRs $A_i$. The cardinality of the KSG is:

\begin{equation}
\varsigma=2^{\sum_{i\in S}log_2(|A_i|)}
\end{equation}
\end{theorem}
This results with cardinality: $\varsigma\approx2^{100}$

Furthermore, the 223 initial state is defined randomly by means of the TRNG adding an entropy of about 223 bits.

%2.1_BCF
\subsection{Boolean combining Function}
The Algebraic Normal Form (ANF) of the proposed Boolean combining function is as follows:

\begin{equation}\label{BCF}
\begin{split}
F(x_1,...,x_{16})=x_1+x_2+x_3+x_4+x_5+x_6+x_7+x_8 \\
+x_9x_{11}+x_{10}x_{11}+x_{10}x_{12}+x_{13}x_{15}+x_{14}x_{15}+x_{14}x_{16} \\
+x_9x_{10}x_{11}+x_{10}x_{11}x_{12}+x_{13}x_{14}x_{15}x_{16}
\end{split}
\end{equation}

\noindent This Boolean combining function consists of two parts:
\begin{itemize}
  \item The linear part, which contains the monomials of degree one $x_1$ to $x_8$, can be realized with two 4-LUTs, 
  \item The non-linear part containing monomials of degree two and three, related to the terms $x_9$ to $x_{16}$ can also be realized with another two 4-LUTs. 
The outputs of all four 4-LUTs are combined with one 4-LUT to generate the keystream $Z_t$. 
\end{itemize}

\paragraph{Properties of the BF. }
\begin{definition}
The Boolean combining function $F$ can be described as follows:
\begin{equation}\label{_BCF_GF}
F: \{0,1\}^{16} \rightarrow \{0,1\}
\end{equation}

\end{definition}

The structure of a Boolean combining function is widely deployed in stream cipher designs. Although LFSR (linear feedback shift register)/ NLFSR (non-linear feedback shift register) can be excellent pseudo-random generators for being efficient, fast and with good statistical properties. They may be prone to be attacked due to linearity. \\
A secure combining Boolean function should have the following properties: balanced, high algebraic degree, high correlation immunity and high nonlinearity. In the following, we present the analysis results of the Boolean combining function $F$.

\subsubsection{Balanced}
A Boolean combining function is balanced if and only if the numbers of ‘1’s and ‘0’s in its truth table are equal.
Since the LFSRs/NLFSRs are supposed to be randomly i.d.d. (independent identically distributed), the combining function must be balanced to satisfy the pseudo-randomness. Otherwise, by inputting a large number of randomly selected values, the output value will not be balanced. Furthermore, the whole system would be vulnerable to cryptanalytic attacks. \\
The Boolean combining function $F$ is balanced.

\subsubsection{Algebraic degree}
The algebraic degree is the degree of ANF of the Boolean combining function. Since the ANF of the Boolean combining function has degree 4, the algebraic degree of $F$ is 4.

\subsubsection{Correlation immunity} 
Before we introduce the correlation immunity, an introduction to Walsh Transformation is needed.

\begin{definition}
Let  $x=(x_1,x_2,...,x_n)$ and  $\omega=(\omega_1,\omega_2,...,\omega_n)$ be n-tuples over $\{0,1\}$  , and dot product of $x$ and  $\omega$ is defined as:

\begin{equation}
x.\omega=x_1\omega_1+x_2\omega_2+...+x_n\omega_n
\end{equation}

\noindent Then the Walsh Transformation on a n-variable Boolean function $f(x)$  is defined as :

\begin{equation}
F(\omega)=\sum_{x}{f(x)(-1)^{x\omega}}
\end{equation}

The Correlation immunity can be calculated based on the Walsh Transformation as follows:
If for all $1\leq wt(\omega) \leq t$, with $wt(\omega)$ is the weight of $\omega$, the Walsh Transformation $F(\omega)=0$ , then the number $t$ is called correlation immunity.  
\end{definition}

\noindent The correlation immunity of the Boolean combining function $F$ is 8.

\subsubsection{Nonlinearity}
The nonlinearity is the distance from the combining function $F$ to the set of affine functions of $n$-variables ($A_n$):

\begin{equation}
NL(F)=min_{h\in A_n} d(F, h)
\end{equation}

\noindent The non-linearity of the combining function $F$ is:   $NL(F)=26624$

\subsubsection{Algebraic immunity} 
For $F: \{0,1\}^m \rightarrow \{0,1\}$, define $AN(F)=\{g: \{0,1\}^m \rightarrow \{0,1\}/F.g=0\}$, any function $g\in AN(F)$ is called the annihilator of $F$. 
The algebraic immunity of $F$ is the minimum degree of all the nonzero annihilators of $F$ and of all those of $F+1$.  In \cite{Courtois2003a}, it was proved that the algebraic immunity is less than or equal to $n/2$ for any $n$-variable Boolean function $F$.
The algebraic immunity of the Boolean combining function $F$ is 4.\\

\noindent As a summary, the Boolean combining function $F$ is balanced with algebraic degree 4, correlation immunity 8, nonlinearity 26624 and algebraic immunity 4.

%3. Security Analysis
\section{Security Analysis}
\label{sec:sec3}
In this section, we present a security analysis of the stream cipher against the following attacks: Brute force attack, Correlation Attack, Algebraic Attack and Parity Check Attack. 

\subsection {Brute force attack}
\subsubsection{Exhaustive search of NLFSRs initial states}
The first brute force attack is exhaustive search of all the internal states in NLFSRs. The Adversary will enumerate all possible states, then generate the corresponding sequence in each case and compare it with the known keystream. If the generated sequence and the keystream match, then the internal states of the NLFSR are found and the cipher is broken.\\
The complexity of this attack is:

\begin{equation}
Complexity=2^{\sum_{i=1}^{i=16}{N_i}+\sum_{i\in S}log_2(|A_i|)}
\end{equation}

The total length of the NLFSRs is 223 bits and the cardinality of the KSG is $\approx 2^{100}$ resulting with a complexity of $\approx 2^{323}$.
We conclude that the complexity of brute force attack to guess all possible internal states is beyond possible computational power. So the cipher is secure against this attack.

\subsubsection{Berlekamp-Massey algorithm}
The other smarter and important brute force attack algorithm is the Berlekamp-Massey algorithm \cite{Massey1969}. In order to analyse the complexity of the B-M algorithm on the proposed cipher, it is necessary to compute lower bound of the total linear complexity of the output bit-stream. The time complexity of the B-M algorithm attack is the square of the total linear complexity, and the data complexity is the double.
If the lengths $N_1,...,N_t$ of the $t$ shift registers are pairwise relatively prime, then the linear complexity $L(\zeta)$ of the keystream $\zeta$ can be expressed as \cite{Rueppel1986}:

\begin{equation}
L(\zeta)\geq F(L_1,...,L_t)
\end{equation}

If the lengths of the primitive NLFSRs are not pairwise relatively prime, then equation (5) does not hold. In this case,  $F(L_1,...,L_t)$ provides only an upper bound for $L(\zeta)$. However, in many cases, it is still possible to derive a reasonable lower bound for the linear complexity of $\zeta$. 

\begin{lemma}\cite{Gammel2006}
Let  $\sigma_1,...,\sigma_t $ be nonzero output sequences of primitive binary NLFSRs of lengths  $N_1,...,N_t$, respectively, and with linear complexities $L_1,...,L_t$ , respectively. Let $F(x_1,...,x_t)$  be a Boolean function of algebraic degree $d\geq1$ . A lower bound for the linear complexity of the sequence $\zeta=F(\sigma_1,...,\sigma_t)$  can be given if the following two conditions are fulfilled:

\begin{enumerate}
\item The algebraic normal form (ANF) of $ F(x_1,...,x_t)$ contains a monomial $x_{i_1},...,x_{i_d}$  of degree $d$ for which the corresponding shift register lengths  $N_1,...,N_d$  are pairwise relatively prime.
\item For all other monomials of degree $d$, which have the form  $x_{i_1},...,x_{i_j-1}x_{i_k}x_{i_j+1},...,x_{i_d}$, we have $gcd(N_{ij},N_k)=1$.
\end{enumerate}

If both conditions are true, then: 

\begin{equation}
L(\zeta)\geq L_{i_1}L_{i_2}...L_{i_d}
\end{equation}
\end{lemma}
The Boolean combining function $F$ has algebraic degree 4, the ANF of it contains the following monomial with degree 4: 
$$x_{13}x_{14}x_{15}x_{16}$$

\begin{itemize}
\item The monomial $x_{13}x_{14}x_{15}x_{16}$ satisfies the condition 1 in Lemma 1: The lengths of the corresponding shift registers are $N_{13}=19, N_{14}=21, N_{15}=22, N_{16}=23$ are pairwise relatively prime. 

\item The other monomials in the ANF of the Boolean combining function are of degree less than the degree of the monomial described in (6). Then condition 2 holds.
\end{itemize}

We conclude that the linear complexity of the keystream $\zeta$ is:

\begin{equation}
L(\zeta)\geq L_{i_{13}}L_{i_{14}}L_{i_{15}}L_{i_{16}}>(2^{18}+19).(2^{20}+21).(2^{21}+22).(2^{22}+23)\approx 2^{81}
\end{equation}

\noindent B-M algorithm requires $2^{162}$ time complexity and $2^{82}$ data complexity to break the proposed KSG.

\subsection {Correlation attack}
The correlation attack was firstly proposed by T. Siegenthaler in 1984 \cite{Siegenthaler1984}, then improved by W. Meier and O. Staffelbach in 1989 as fast correlation attack \cite{Meier1989}. The main idea of correlation attack is to focus on the Boolean combining function of the KSG and find the correlation between the combination of several LFSRs/NLFSRs and the output keystream. This requires to have previous knowledge about the used NLFSRs, i.e. an adversary should reverse the randomly selected feedback functions before applying this attack. Since, there is more than $2^{100}$ possible combinations of feedback functions and about $2^{223}$ initial states, trying to reverse the feedback functions is not possible.

Now, let us consider that an adversary knows the used feedback functions. In this case, the adversary can apply correlation attack with an aim to recover the NLFSRs initial states. By considering the classical fast-correlation attack, when the Boolean combining function has correlation immunity $n$, the adversary needs at least $n+1$ shift registers at the same time. 
The correlation immunity of the Boolean combining function $F$ is 8; therefore the total length of the smallest 9 NLFSRs is:

\begin{equation}
\sum_{i=6}^{i=14}{i}=90
\end{equation}

\noindent Thus, if an adversary is given the used feedback functions of an SUC, the complexity of the correlation attack is at least $2^{90}$. However, this attack can not be realized since for each SUC the random feedback functions are selected randomly and internally inside the chip.

\subsection {Algebraic attack }
The algebraic attack is another important attack against stream ciphers. It was introduced in \cite{Courtois2003} and extended in \cite{Courtois2003a}. The main idea of the Algebraic attack is to find a well-chosen multivariate polynomial $G(s)$, such that $G.F$ is of substantially lower degree, where $F(s)$ is the combining Boolean function and $s$ is the current state.
To examine the degree of the linear polynomial equations system, an assertion for the degree of the algebraic equations from \cite{Gammel2006a} will be used. It is described in the following fact:
\paragraph{Fact \cite{Gammel2006a}}
For $2N_j\leq2^{N_j}-N_j$, the $k^{th}$ entry in the monomial spectrum of the shift registers $A_j$, with  $1\leq j \leq16$, contains close to   $2^{N_{j-1}}$ different monomials and has in general degree $N_{j-1}$. 

For the Boolean combining function $F$, we find the term with highest degree and then calculate it as follows:
The highest term is $x_{13}x_{14}x_{15}x_{16}$  and its degree can be calculated as:

\begin{equation}
(N_{13}-1)(N_{14}-1)+(N_{15}-1)+(N_{16}-1)=(19-1)+(21-1)+(22-1)(23-1)=81
\end{equation}

Since we need $2^{N_j}-2$  different monomials, in order to express the bits of the sequence by the initial state of each register. So we need:

\begin{equation}
(2^{N_{13}}-2)(2^{N_{14}}-2)(2^{N_{15}}-2)(2^{N_{16}}-2)\approx2^{81}
\end{equation}

Different monomials in order to express the bits of the sequence from the highest degree term.  Set apart the remaining different monomials, the complexity for solving the system of equations is:

\begin{equation}
O((2^{81})^\omega)=O(2^{192.78})
\end{equation}

Where $\omega \approx 2.38$ is the exponent of the fast matrix multiplication.
The complexity of solving the system of equations ensures that the proposed algorithm is secure against algebraic attack.

\subsection {Parity Check attack }
The parity check attack was firstly proposed in \cite{T.Johansson2006}, it can successfully break the Achterbahn stream cipher. It starts with the weakness of the Boolean combining function that if two terms are equal to 0, then the whole nonlinear part would be 0, therefore the Boolean combining function is purely linear. After the linearization of the Boolean combining function, a parity check is applied in order to retrieve possible inner states of some certain registers. \\

Parity check attack is very sensitive to the number of terms in the combining function after linearization, because that  bits/terms are required to complete the parity check and the number also determines the expected number of the inner states in certain registers, which satisfy the precondition of the linearisation (some bits are equal to 0).\\
In the following, we present the security analysis of the proposed algorithm against parity check attack. 
The ANF of the Boolean combining function $F$ is defined as:

\begin{equation}\label{BCF}
\begin{split}
F(x_1,...,x_{16})=x_1+x_2+x_3+x_4+x_5+x_6+x_7+x_8 \\
+x_9x_{11}+x_{10}x_{11}+x_{10}x_{12}+x_{13}x_{15}+x_{14}x_{15}+x_{14}x_{16} \\
+x_9x_{10}x_{11}+x_{10}x_{11}x_{12}+x_{13}x_{14}x_{15}x_{16}
\end{split}
\end{equation}

It has a linear part and a nonlinear part. When we examine the common terms of the nonlinear part, if $x_9=x_{10}=x_{13}=x_{14}=0$, then the Boolean combining function would degenerate into a pure linear Boolean combining function:

\begin{equation}
l(x_1,...,x_{16})=x_1+x_2+x_3+x_4+x_5+x_6+x_7+x_8
\end{equation}

The upper bound of the linear complexity is relatively not that high:

\begin{equation}
L\geq \sum_{i=1}^{i=8}{L_i} \approx 2^{14.9}
\end{equation}

So we can build an LFSR with length $L$, and we can also apply parity check on the sequence output of the sequence from the linear Boolean combining function.

\paragraph {Parity Check.} The periods of all the NLFSRs are as follows:

\begin{equation}\label{BCF}
\begin{split}
T_1=2^6-1; T_2=2^7-1; T_3=2^8-1; T_4=2^9-1; T_5=2^{10}-1; \\
T_6=2^{11}-1; T_7=2^{12}-1;  T_8=2^{13}-1; 
\end{split}
\end{equation}

\noindent Where $T_i$  denotes the period of the NLFSR $A_i$. \\
\\
\noindent So, the parity check can be computed as follows:

\begin{equation}
ll(t)=l(t)\oplus l(t+T_1)
\end{equation}

\noindent Since the period of the first register is $T_1$, this expression does not contain any term of $x_1$.\\
\\

\noindent Similarly:

\begin{equation}
\begin{aligned}
\begin{split}
lll(t)&=ll(t)\oplus ll(t+T_2)\\
llll(t)&=lll(t)\oplus lll(t+T_3)\\
lllll(t)&=llll(t)\oplus llll(t+T_4)\\
llllll(t)&=lllll(t)\oplus lllll(t+T_5)\\
lllllll(t)&=llllll(t)\oplus llllll(t+T_6)\\
llllllll(t)&=lllllll(t)\oplus lllllll(t+T_7)\\
\end{split}
\end{aligned}
\end{equation}

\noindent Therefore $llllllll(t)$ contains only terms of $x_8$. Thus it satisfies:
$$llllllll(t)=lllllll(t)\oplus lllllll(t+T_8)$$

\noindent In terms of bits $l(i)$ , we have the following equation:

%\begin{equation}
%\begin{split}
\begin{align*}
& l(t)+l(t+T_1)+l(t+T_2)+l(t+T_3)+l(t+T_4)+l(t+T_5)+l(t+T_6)+l(t+T_7)+\\
& l(t+T_8)+l(t+T_1+T_2)+...+l(t+T_1+T_8)+l(t+T_2+T_3)+...+l(t+T_2+T_8)+...\\
& +l(t+T_1+T_2+T_3)+...+l(t+T_1+T_2+T_8)+...\\
& +...\\
& +l(t+T_1+T_2+T_3+T_4+T_5+T_6+T_7+T_8)=0
%\end{split}
%\end{equation}
\end{align*}

This is the basic parity check on $l(t)$  that can be used to attack the KSG, it is the XOR between 256 bits from the sequence , within the time interval:
$$ T_{max}=T_1+T_2+T_3+T_4+T_5+T_6+T_7+T_8 \approx 2^{15}$$

It is the complexity required for the parity check attack, but the degeneration happens under the condition $x_9=x_{10}=x_{13}=x_{14}=0$ and the complexity to satisfy this condition should be considered.

Consider the  $x_9$  register first, which   $x_9=0$ is the condition for further parity check attack. For every bit used in parity check, totally 256 bits, they should all satisfy the condition. The number of all the possible internal states in register  $x_9$  is $2^{14}$. Since the output should be independent and identically distributed (i.d.d), the expected number of cases that satisfy this condition is:

\begin{equation}
2^{14}\times2^{-256}=2^{-242}
\end{equation}

At this step, the attack cannot continue, since the possibility of finding a case satisfying the condition  $x_9=0$  is too small to ignore.

%4
\section{Application of Secret Unknown Ciphers}
\label{sec:sec4}
%\subsection{General architecture}
Secret Unknown Cipher (SUC) is a digital clone-resistant unit that can be deployed as a security anchor in wide spectrum of applications such as automotive security \cite{Mars_V2X}. SUC is a random cipher created internally in SoC FPGAs resulting with unique and unpredictable random cipher even for the producer for each SoC unit. To create SUC, ciphers/hashes with flexible components that can be generated randomly should be designed. In this paper, SUC based on combining random NLFSRs with identical cryptographic properties is presented. Recently, all published  generic protocols deploying SUC \cite{Adi2017e} are designed to be used for SUC based on random block ciphers. Protocols targeted for random stream ciphers or random key stream generators based SUC will be investigated in the following.

%\begin{center}
%\begin{figure}[h]
%\includegraphics[scale=0.55]{SUC_App.pdf}
%\caption{General description of SUC and its application in devices identification}
%\label{fig:SUC_App}
%\end{figure}
%\end{center}

%\subsection{Protocols}

\subsection{Enrollment protocol}
During the enrollment process, the Trusted Authority (TA) stimulates a unit A by a command (cmd) to generate a $k$-bit response $Y$. The $k$-bit response results from running the KSG $k$-cycle. This operation is to be done sequentially to generate all the responses $Y_0,…,Y_{t-1}$.

\begin{center}
\begin{figure}[!htb]
\includegraphics[scale=0.5]{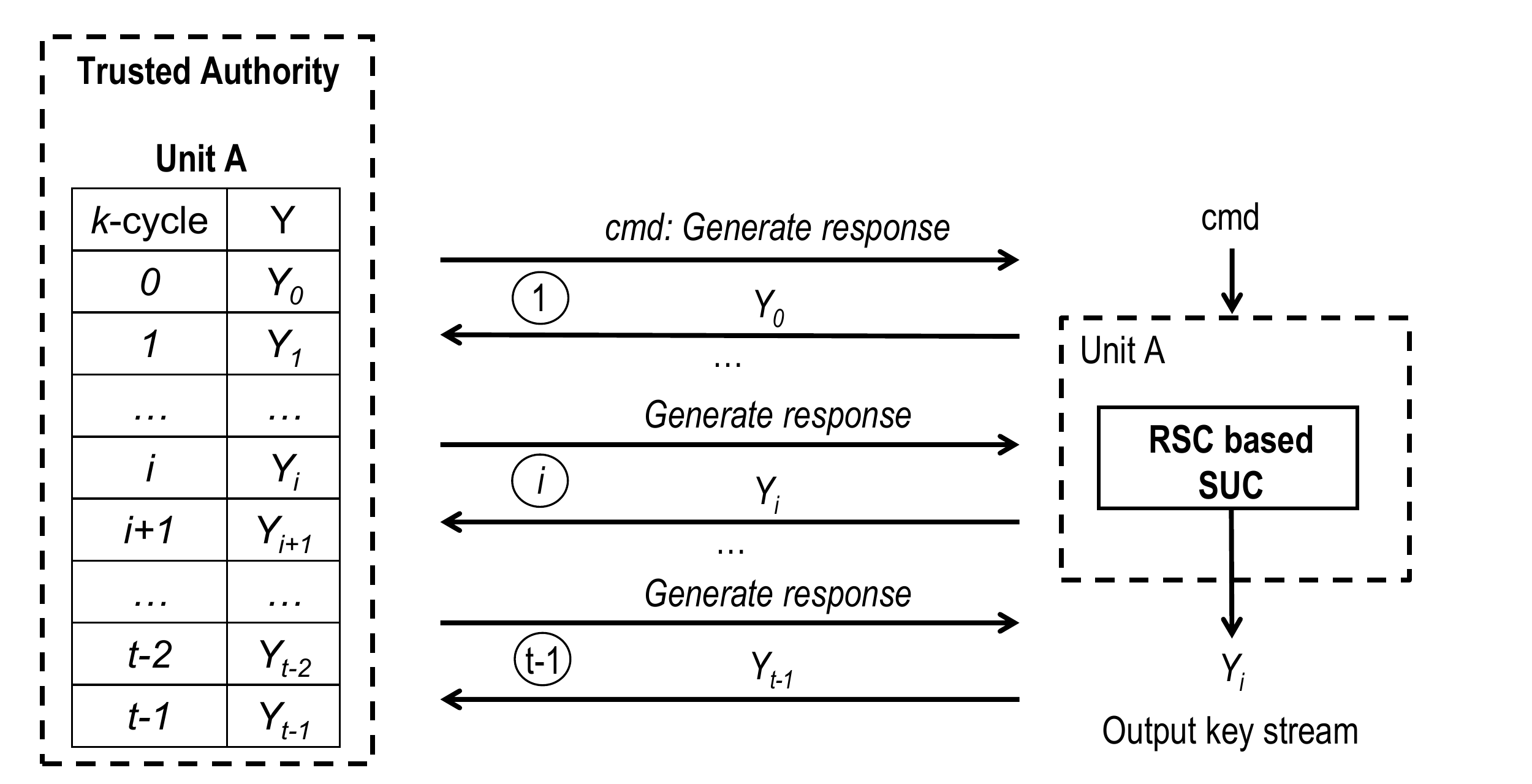}
\caption{Enrollment protocol of SUC based on random stream ciphers}
\label{fig:enrolment}
\end{figure}
\end{center}

\noindent Figure~\ref{fig:enrolment} describes the enrollment protocol.

\subsection{Identification protocol}

\begin{center}
\begin{figure}[!htb]
\includegraphics[scale=0.5]{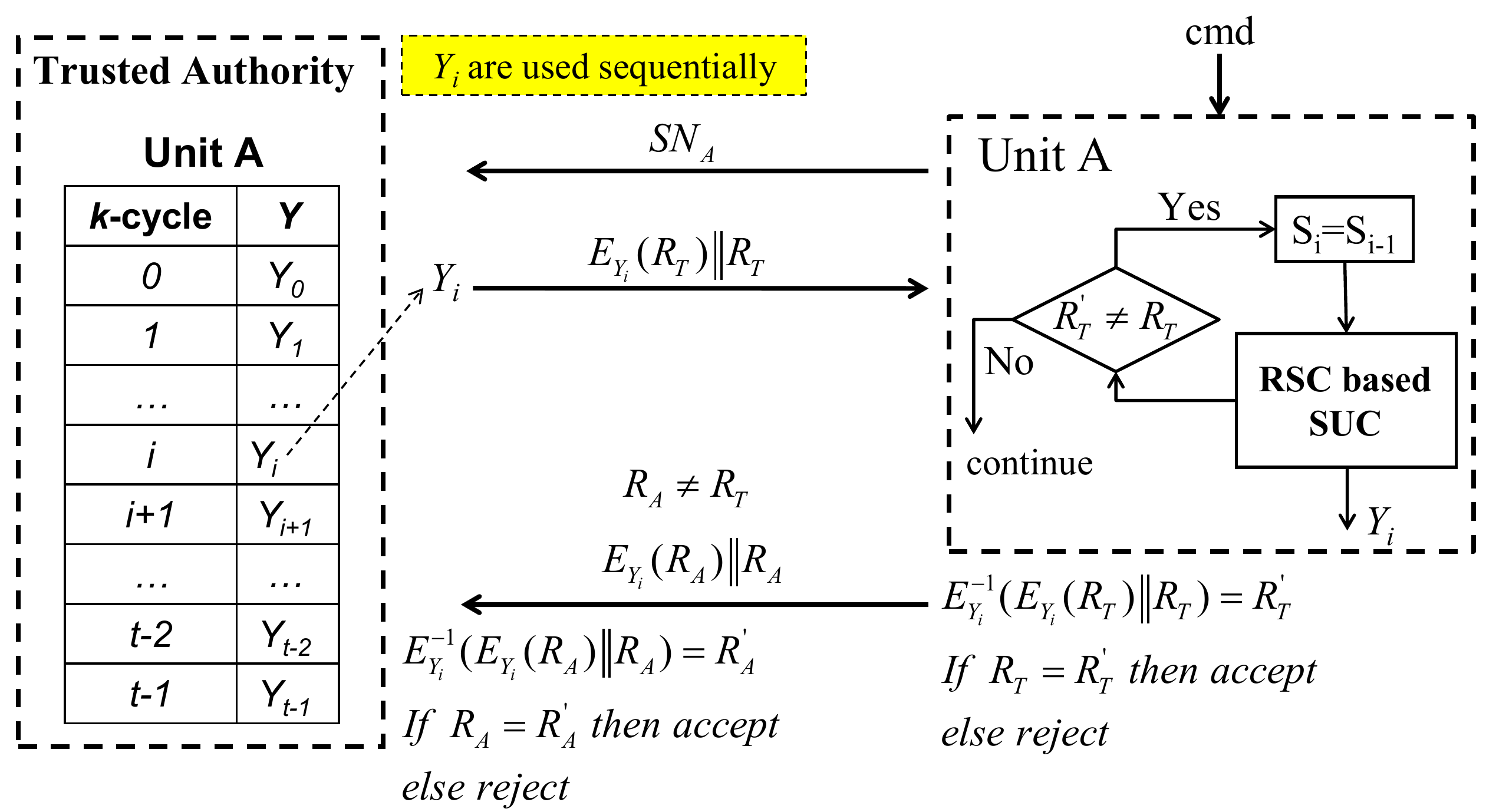}
\caption{Identification protocol of SUC based on random stream ciphers}
\label{fig:identification}
\end{figure}
\end{center}

\noindent Figure~\ref{fig:identification} describes the identification protocol, it proceeds as follows:

\begin{itemize}
\item Unit A sends its serial number $SN_A$ to the TA that checks for its existence in the TA units identification records UIR. if $SN_A\in UIR$, then TA accepts and continues otherwise it rejects and abort.
\item The TA selects the exact $Y_i$ and generates a random number $R_T$. The, it encrypts $R_T$ with a standard cipher by using $Y_i$ as key and sends it in concatenation with $R_T$ as $E_T(R_T)||R_T$.
\item Unit A generates the next response $Y'_i$ and decrypts the received message as $E^{-1}(E_{Y_i}(R_T))=R'_T$. If $R_T \neq R'_T$ then the received message doesn’t come from the TA and the unit rejects and abort the communication. Meanwhile, unit A keeps the state $S_{i-1}$. This protects the system from being desynchronized. When $R_T=R'_T$, unit A generates a random number $R_A$, encrypts it by the same $Y_i$ and sends it concatenated with $R_A$ to TA.
\item TA decrypts the message as $E^{-1}(E_{Y_i}(R_A))=R'_A$. If $R_A=R'_A$, the TA accepts otherwise it rejects and aborts.
\end{itemize}

\subsection{Update protocol}
\begin{center}
\begin{figure}[!htb]
\includegraphics[scale=0.5]{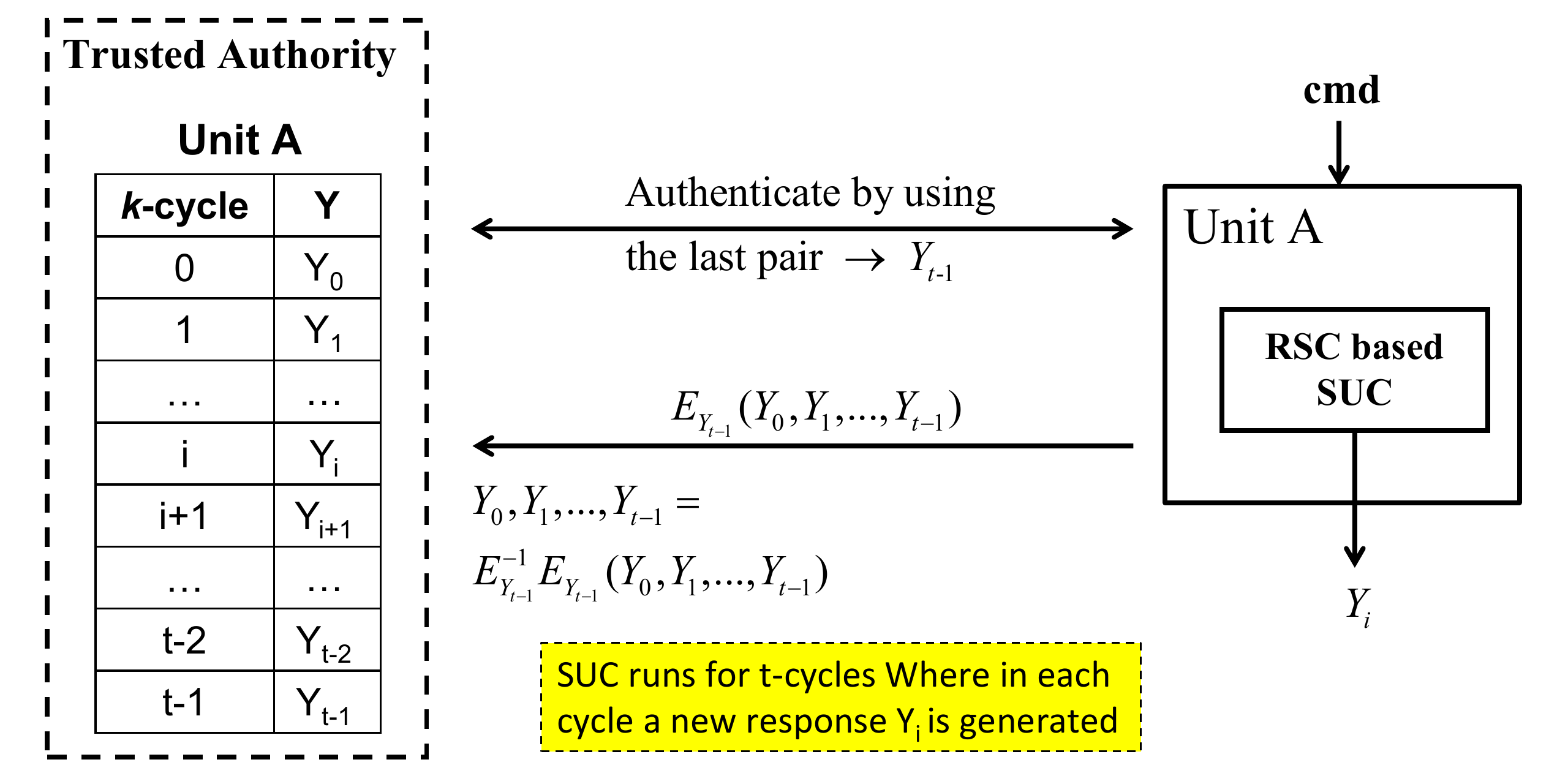}
\caption{Update protocol of SUC based on random stream ciphers}
\label{fig:update}
\end{figure}
\end{center}

\noindent Figure~\ref{fig:update} descibes the update protocol, it proceeds as follows:

\begin{itemize}
\item The TA and  $unit_A$ authenticates each other by using the last response $Y_{t-1}$.
\item  Unit A generates t-responses and sends them encrypted with $Y_{t-1}$ to the TA as $E_{Y_{t-1}}(Y_0, Y_1, ..., Y_{t-1})$.
\item TA decrypts the received data by using $Y_{t-1}$ as $E^{-1}_{Y_{t-1}}(E_{Y_{t-1}}(Y_0, Y_1, ..., Y_{t-1}))=Y_0, Y_1, ..., Y_{t-1}$ and updates unit A responses in its UIR.
\end{itemize}

%\subsection{Cardinality of the KSG}
%The proposed KSGs can be used to create a family of SUC. The NLFSRs selected for the proposed design can be made random, since for each NLFSR length there exist a number of possible feedback functions that will be described in the following section. For each length, each feedback function in \cite{Dubrovaa} ensures that the resulting NLFSR has a maximum period and hence randomly selecting one of the feedback functions for each NLFSR $A_i$ will ensure the same security level of the resulting random KSG as the one that was investigated through this paper.
%
%In \cite{Dubrovaa}, a list of NLFSRs with maximum period  $2^{N_i}-1$, of length $3<N_i <25$ has been presented. Table presents the number of NLFSRs for each length described in \cite{Dubrovaa}:
%
%\begin{table}[H]
%\begin{center}
%
%\input{All_NLFSRs.tex}
%\caption{Number of all listed NLFSRs as listed in \cite{Dubrovaa}}
%\vspace{1ex}
%\end{center}
%\end{table}
%
%The proposed KSG deploys 16 NLFSRs with lengths from 6 to 17, 19, 21, 23 and 24.
%
%
%\begin{theorem}
%Let $N_i$ be the lengths of the NLFSRs of the KSG, where  $i\in S$ where $S=\{6,...,17,19,21,23,24\}$.\\
%We denote by  $|A_i|$the number of NLFSRs $A_i$ of length $N_i$. The cardinality of the KSG is:
%
%\begin{equation}
%\varsigma=2^{\sum_{i\in S}log_2(|A_i|)}
%\end{equation}
%\end{theorem}
%This results with cardinality:     $\varsigma=2^{65.76}$
%
%Furthermore, the 225 initial state is defined randomly by means of the TRNG adding an entropy of about 225 bits.

%5
\section{Hardware Complexity}
\label{sec:sec5}
Mass production requires lightweight identification mechanisms for economic reasons. SUCs as clone-resistant identities gain special interest because it is possible to be implemented with zero-cost. Most FPGA applications do not consume the total resources offered by the deployed FPGA. In such cases, our proposed SUC unit requires very few FPGA resources and can be added incrementally to the existing application for zero-costs.\\
The KSG described in Figure~\ref{fig:KSG} is modeled in VHDL and synthesized to check its hardware complexity and performance. Libero SoC with its integrated tools is used to implement the design; Mentor Graphics Modelsim ME design tool was used for simulation and Synplify pro ME for synthesis. \\
\noindent Table~\ref{table:hwcomplexity} describes the resources consumed for different devices of SmartFusion®2 SoC FPGAs family.

\begin{table}[h]
\begin{center}
\caption{Hardware complexity of the KSG in SmartFusion®2 SoC FPGAs}
\label{table:hwcomplexity}

\begin{tabular}{|c| c| c| c|c|c| c|c|}

\hline
\multicolumn {2} {|c|}{KSG components}& 
\multicolumn {2} {|c|}{Resources usage}&
\multicolumn {2} {|c|}{\thead{\% of usage\\for M2S005}}&
\multicolumn {2} {|c|}{\thead{\% of usage\\ for M2S150}}\\

 \hline
\multicolumn {2} {|c|}{Type of resources}	 &LUTs	&DFFs 	&LUTs 	&DFFs 	&LUTs 	&DFFs\\ 
\hline
NLFSRs          &Shift registers	                     &0	          &223	          &0		&3.71		 &0		&0.15 \\ 
\hline
   	          &Feedback Functions	          &32	          &0	          &0.52   	&0	           &0.02   	&0\\

\hline
\multicolumn {2} {|c|}{Combining Function}    & 5	          &0	         &0.09		&0	          &0.005	&0\\
\hline
\multicolumn {2} {|c|}{Total}                           &37	          &223	         &0.61  	         &3.71		&0.025	&0.15\\
\hline

\end{tabular}
\vspace{1ex}
\end{center}
\end{table}

\paragraph {}
\noindent The KSG requires 37 LUTs and 223 DFFs, this can be considered as zero cost in many real applications. Implementation mechanism of the SUC based on the proposed family of new stream ciphers is not presented in this paper, a concept for creating SUC in SmartFusion®2 SoC FPGAs is presented in \cite{Mars2017}.

\paragraph{Implementation method}

\begin{figure}[H]
\begin{center}
\includegraphics[scale=0.55]{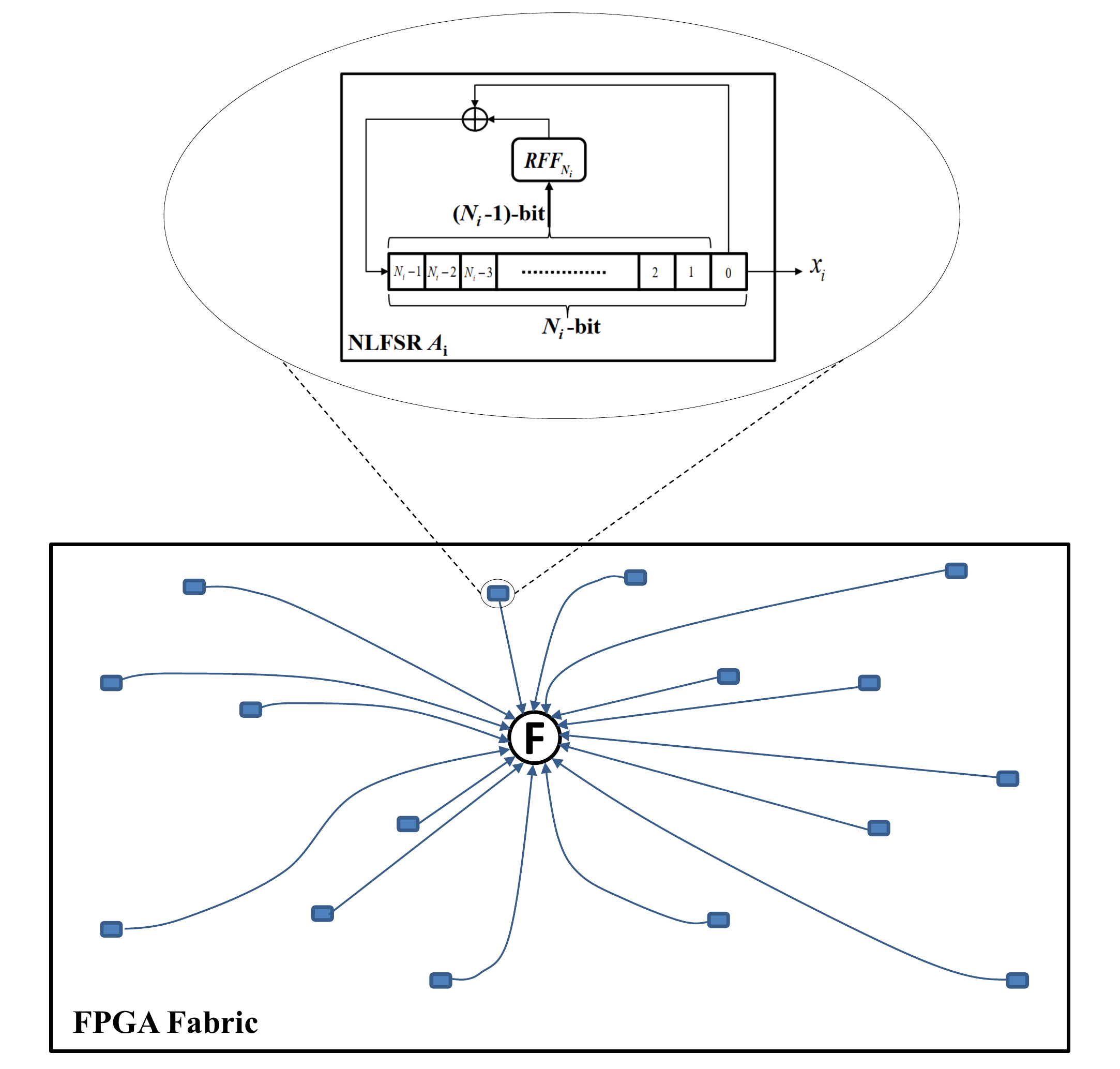}
\caption{Combining random NLFSRs to build an SUC}
\label{fig:sc_hw}
\end{center}
\end{figure}

Figure~\ref{fig:sc_hw} describes an FPGA implementation concept of SUC based on combining random NLFSRs. Each NLFSR is providing one bit per cycle to the combining function. This makes it possible to distribute NLFSRs over all the FPGA where only one connection is required to connect an NLFSR to the combining function. This constitutes an advantage over SUC based on block cipher designs.

%6
\section{Conclusion}
\label{sec:sec6}
In this paper, a new large class of Key Stream Generators KSGs as stream ciphers is presented. The class is created by random selection of a set of maximum-period NLFSRs with different lengths. It is shown that a random selection of one Secret Unknown Cipher (SUC) from this class, can serve to convert future VLSI-devices (in a post-fabrication process) into clone-resistant entities. The security level of the proposed cipher class against many attacks is evaluated and shown to be scalable to cope even with post-quantum security requirements. i.e. as the attack complexity exceeds $2^{160}$, the cipher structure can cope even with post-quantum cryptographic requirements at relatively moderate implementation complexity. A lightweight proposed generic Identification/authentication protocol for such physical SUC based structures is also presented. A sample prototype case showed that one SUC structure consumes relatively minor FPGA resources; (0.61\% of the LUTs, 3.71\% of DFFs) in the smallest Microsemi SmartFusion®2 SoC FPGA M2S005 devices, and (0.025\% LUTs, 0.15\% DFFs) for the largest M2S150 device. Future work is in progress to fine-tune and optimize ciphers and SUC structures for emerging VLSI technologies.

%Acknowldgment
\section*{Acknowldgment}
This work was supported by Volkswagen AG and Microsemi, a Microchip Company, San Jose USA as well as the German Federal Foreign Office funding by DAAD combined scholarship and support program (STIBET).
%\section{Appendix: new version of the table}
%\input{All_NLFSRs_table_v2.tex}
%References
\bibliographystyle{alpha} 
\bibliography{TCHES_refs}

\end{document}